\documentstyle[11pt,epsfig]{article}                                           

\newcommand{\dgpicture}[2]{
\begin{picture}(#1,#2)   
\thicklines
}

\newcommand{\epsfigdg}[2]{\epsfig{figure=#1,#2}}
   
\newcommand{\fullline}{                                                        
\unitlength0.4mm                                                               
\begin{picture}(13,4)                                                          
\linethickness{0.3mm}                                                          
\put(-1,2.0){\line(1,0){15}}                                                   
\thinlines                                                                     
\end{picture}                                                                  
}                                                                              
\newcommand{\dashline}{                                                        
\unitlength0.4mm                                                               
\begin{picture}(20,4)                                                          
\linethickness{0.3mm}                                                          
\put(-1,2.0){\line(1,0){4}}                                                    
\put(8,2.0){\line(1,0){4}}                                                     
\put(17,2.0){\line(1,0){4}}                                                    
\thinlines                                                                     
\end{picture}                                                                  
}                                                                              
\newcommand{\dotline}{                                                         
\unitlength0.4mm                                                               
\begin{picture}(9,4)                                                           
\linethickness{0.3mm}                                                          
\put(-1,2.0){\line(1,0){1}}                                                    
\put(4,2.0){\line(1,0){1}}                                                     
\put(9,2.0){\line(1,0){1}}                                                     
\thinlines                                                                     
\end{picture}                                                                 
}                                                                              
\newcommand{\longdashline}{                                                    
\unitlength0.4mm                                                               
\begin{picture}(22,4)                                                          
\linethickness{0.3mm}                                                         
\put(-1,2.0){\line(1,0){10}}                                                   
\put(13,2.0){\line(1,0){10}}                                                   
\thinlines                                                                     
\end{picture}                                                                  
}                                                                              
\newcommand{\dashdotline}{                                                     
\unitlength0.4mm                                                               
\begin{picture}(17,4)                                                          
\linethickness{0.3mm}                                                          
\put(-1,2.0){\line(1,0){5}}                                                    
\put(8,2.0){\line(1,0){1}}                                                     
\put(13,2.0){\line(1,0){5}}                                                    
\thinlines                                                                     
\end{picture}                                                                  
}                                                                              
                                                                              
\newcommand{\dashdotdotline}{                                                  
\unitlength0.4mm                                                               
\begin{picture}(11,4)                                                          
\linethickness{0.3mm}                                                          
\put(-1,2.0){\line(1,0){5}}                                                    
\put(8,2.0){\line(1,0){1}}                                                     
\put(11,2.0){\line(1,0){1}}                                                    
\thinlines                                                                     
\end{picture}                                                                  
}                                                                              
\newcommand{\dashdashdotdotline}{                                              
\unitlength0.4mm                                                               
\begin{picture}(22,4)                                                          
\linethickness{0.3mm}                                                          
\put(-1,2.0){\line(1,0){5}}                                                    
\put(9,2.0){\line(1,0){5}}                                                     
\put(19,2.0){\line(1,0){1}}                                                    
\put(22,2.0){\line(1,0){1}}                                                    
\thinlines                                                                     
\end{picture}                                                                  
}

\newcommand{\be}{\begin{equation}}                                             
\newcommand{\ee}{\end{equation}}                                               
\newcommand{\beqn}{\begin{eqnarray}}                                           
\newcommand{\eeqn}{\end{eqnarray}}                                             
\newcommand{\spc}[1]{\mbox{\hspace{#1}}}                                       
\begin{document}                                                               
\begin{titlepage}                                                              
\hfill
\hspace*{\fill}
\begin{minipage}[t]{4cm}
DESY--96--36\\
CERN--TH/96--83\\
ANL-HEP-PR-96-23\\
EDINBURGH-96-2
\end{minipage}
\vspace*{2.cm}                                                                 
\begin{center}                                                                 
\begin{LARGE}                                                                  
{\bf Associated Jet Production at HERA}\\                                      
\end{LARGE}                                                                    
\vspace{2cm}                                                                   
{\bf J.~Bartels$^1$, V.~Del Duca$^2$, A.~De Roeck$^3$, D.~Graudenz$^4$,
M.~W\"usthoff$^{5,a}$ 
}\\                                                    
\end{center}                                                                   
$^1$ II.~Institut f\"{u}r Theoretische Physik, Universit\"{a}t Hamburg,        
D-22761 Hamburg, Germany.\\                                                    
$^2$ Dept. of Physics and Astronomy, University of Edinburgh,  
Edinburgh EH9 3JZ, Scotland,
U.K.\\                                                             
$^3$ Deutsches Elektronen Synchrotron DESY, D-22603 Hamburg, Germany.\\        
$^4$ CERN, Theory Division, 1211 Geneva 23, Switzerland.\\                     
$^5$ Argonne National Laboratory, 9700 South Cass Avenue, IL 60439, 
USA.\\ 
$^a$ Partly supported by the U.S. Department of Energy, 
                            Division of High Energy Physics,
                            Contract W-31-109-ENG-38. 
\\
\vspace*{2.cm}                          
\begin{center}                                       
{\bf Abstract}
\end{center}                                                                
\begin{quotation}                                                              
   We compare the
  BFKL prediction for the associated production of forward jets at HERA 
  with  fixed-order matrix element calculations 
  taking into account the kinematical cuts imposed by experimental
  conditions. Comparison with H1 data of the 1993 run favours
  the BFKL prediction. As a further signal of BFKL dynamics, we propose to
  look for the azimuthal dependence of the forward jets. 
\end{quotation}                                                                
\vfill                                                                         
\vfill
\noindent
\vspace{1cm}
\end{titlepage}                                                                
\newpage                                                                       
\vspace{2cm}                                                                   
\noindent                                                                      
{\bf 1.~Introduction:}                                                          
The BFKL \cite{BFKL} Pomeron has recently gained much interest, in            
particular in connection with the rise of $F_2$ at small 
Bjorken-$x$ ($x_B$) observed at       
HERA. Since it is generally believed that, at small $x$,  $F_2$ is driven
by the gluon density in the proton, it is tempting to use the BFKL evolution   
equation for calculating the input distribution $g(x, Q_0^2)$.                 
However, such quantitative BFKL-based predictions for $xg(x, Q^2)$ are         
plagued by theoretical difficulties, and it is therefore important             
to look for other signals of the BFKL dynamics. 
A proposed clean                        
``BFKL footprint'' at HERA is the measurement of inclusive                     
jets in deep inelastic scattering                                              
in the forward direction \cite{Mue} with longitudinal momentum                
close to the proton, i.e. the fractional momentum of the jet 
$x_{jet} \gg x_B$,
 and transverse momentum, $k_t$, of the order of the virtual photon 
mass $Q^2$. 
Analytic calculations \cite{Hot} and recent data from H1                
\cite{H1} show encouraging agreement, but the comparison                      
still suffers from several shortcomings. First, the existing theoretical       
estimates of the cross section do not take into account the full experimental  
cuts, which are actually used in the data analysis. Secondly, in order to      
provide a solid ``proof of existence" of the BFKL Pomeron,                     
it is necessary to compare the data not only with                              
the BFKL calculation but also with the fixed-order jet production cross
section based upon QCD matrix elements. The BFKL prediction --- which, 
as a result of gluon production between the forward jet and the current        
jet, rises as a function of $1/x_B$ --- is expected to
lie above the matrix element calculation, and the data will discriminate       
between the two predictions. A third obstacle in verifying the                 
BFKL prediction is the lack of hadronization in the theoretical                
analysis: Monte Carlo studies indicate that, for this type of measurements,    
hadronization may not be  too strong an effect, but there does remain a    
serious uncertainty. An improvement in this direction might be achieved by the 
development of a BFKL-based Monte Carlo program. 
Further evidence for the BFKL Pomeron can be obtained by looking into          
the azimuthal dependence of the forward jet cross section: in the              
HERA reference frame BFKL dynamics predicts \cite{BDW} that, with             
increasing rapidity difference between the forward jet and the current         
jet, the forward jet ``forgets" about the azimuthal direction defined         
by the outgoing electron, and the cross section becomes                        
$\Phi$-independent. This decorrelation is, again, a result of the              
radiation of gluons                                                            
between the forward jet and the current jet. Correspondingly, in               
the fixed-order matrix element calculation one expects the $\Phi$-dependence 
to persist, even for large rapidity intervals.                      
For the Mueller--Navelet \cite{MN} jet production at hadron colliders          
this decorrelation effect has been investigated \cite{D}, and it has         
been demonstrated that such a measurement might give a clear signal for        
the BFKL Pomeron between the two jets. For the analysis of experimental data
from HERA                
an analytic calculation was performed recently \cite{BDW}.               
A numerical analysis, however, has not been done yet. \\ \\                    
In the present paper we will first improve on the comparison between           
the cross section formula and the data. Based upon                             
the analytic expressions given in Ref.~\cite{Hot} we compute the cross
section, taking into account the full experimental cuts used in the recent     
H1 analysis \cite{H1}. We then make use of the fixed-order matrix element     
calculations for the two- and three-jet cross section \cite{GM91} and         
compute, with the same phase space cuts, the numerical                         
predictions. Both calculations are then compared with                          
the data. In the final part of this paper we                                   
suggest a new measurement of the azimuthal decorrelation                       
to reveal the BFKL Pomeron. We                                                 
discuss and present a numerical                                                
analysis of the decorrelation in the azimuthal angle $\Phi$, both              
in the BFKL framework and within the fixed-order jet cross section             
calculation.\\ \\                                                              
{\bf 2.~Experimental Considerations:}                                           
The study of associated jet production is an experimental challenge.           
It was shown in Ref.~\cite{Hot} that the requirement of 
$x_{jet}/x_B$ to be large results in typical jet angles of a few
degrees with respect to the proton direction, also termed forward              
direction. Due to the unavoidable                                              
beam-pipe hole for the detectors at the $ep$ collider HERA, the                 
acceptance is limited to jets with an angle larger than, for example,
$6^\circ$
in               
the present H1 detector.  At smaller  angles the jets are                
insufficiently contained in the detector and the experimental separation      
from proton remnant  fragments is difficult.                                   
The angular requirement leads to maximally reachable                           
$x_{jet}$-values of about 0.1.                                                 
Furthermore, due to the limited amount of data, a compromise has            
to be found for the condition $k^2_t/Q^2 \sim 1$. This condition is essential  
since it defines a kinematical environment where no ambiguity between the      
DGLAP evolution and the BFKL Pomeron exists.                                   
In practice an interval is defined around this central value of $k_t^2/Q^2$.   
The H1 collaboration has made a first measurement of the associated            
or forward-jet production in deep inelastic scattering \cite{H1}.              
The selections are guided by the limited statistics (300 $\mbox{nb}^{-1}$)
and can be refined for future large-statistics samples.                        
In this paper we  adopted similar selection criteria.                          
Deep inelastic events are selected by requiring the energy and angle           
of the scattered electron to be larger than 12 GeV and less than
$173^\circ$,
 respectively.   
The first requirement avoids regions with  large contributions of              
radiative DIS events and experimental background                               
due to misidentified photoproduction events, while the second                  
matches the acceptance of the  calorimeter that detects the scattered 
electron.                            
In order not to confuse the forward jet with the one at the top of the ladder,  
the requirement $y > 0.1$ was imposed                                          
to ensure that the jet of the                                                  
struck quark is well within the central region of the detector                 
and is expected to have a jet angle larger than $ 60^\circ$.
Experimentally                                                                 
a cone algorithm is used to find jets, requiring an $E_t$ larger than          
5~GeV in a cone of radius                                                      
$ R = \sqrt{\Delta\eta^2 + \Delta\phi^2} = 1.0$ in the space of                
 pseudo-rapidity $\eta$ and azimuthal angle $\Phi$                             
in the HERA frame of reference.                                                
For the jet selection we consider two sets of cuts, the first set              
(selection A) matches                                                          
the published data, while the second set                                       
(selection B) could be used when more statistics                               
becomes available and the systematics gets under better control.               
 For selection A                                                               
jets are accepted as forward jets if                                           
$x_{jet}>0.025$, $0.5< p_{jet}^2/Q^2<4$,                                       
$6^\circ < \theta_{jet}< 20^\circ$ and $p_{jet}> 5 $ GeV,
where $\theta_{jet}$ is the forward jet angle,                                  
$p_{jet}$ is the transverse momentum of the jet, in approximation              
of $k_t$.          Selection B                                                 
 accepts   forward jets if                                                     
$x_{jet}>0.035$, $0.5< p_{jet}^2/Q^2<2$,                                       
$7^\circ < \theta_{jet}< 20^\circ$ and $p_{jet}>3.5 $ GeV.
These selection criteria allow a study of the cross section of forward 
jet production in the region                                                   
$Q^2 \approx 20~{\rm GeV}^2$ and                                               
$1\!\times\!10^{-4}\!<\!x_B\!<\!2\!\times\!10^{-3}$.
Hence the ratio $x_{jet}/x_B$ is always larger than 10.\\ \\  
{\bf 3.~The BFKL Approach:}                                                    
The BFKL approach makes a prediction for the high-energy ($x_B \ll x_{jet}$)
behaviour of the cross section for the process $g \to g + \,(n\, g)\, +        
q\bar{q}$ or $q \to q + \,(n\, g)\, + q \bar{q}$. Consequently,                
the first nonvanishing contributions appear in the                             
(3+1)-jet matrix elements\footnote{The 
notation ``$(n+1)$'' stands for $n$ jets in the current            
fragmentation region and one jet in the target fragmentation region,           
the latter 
consisting of the proton remnants.} arising from graphs with one gluon         
in the $t$-channel (for the (2+1)-jet processes $q \rightarrow qg$ and         
$g \rightarrow q\bar{q}$, there are no contributions with gluons in the       
$t$-channel, and therefore their dependence on $x_{jet}/x_B$ is                 
power-suppressed). A generic diagram is shown in Fig.~1a.                      

\begin{figure}[htbp]      \unitlength 1mm
\begin{center}
\dgpicture{159}{65} 
\put(45,0){\rm (a)}
\put(95,0){\rm (b)}
\put(38,10){
     \leavevmode                                                        
       \input{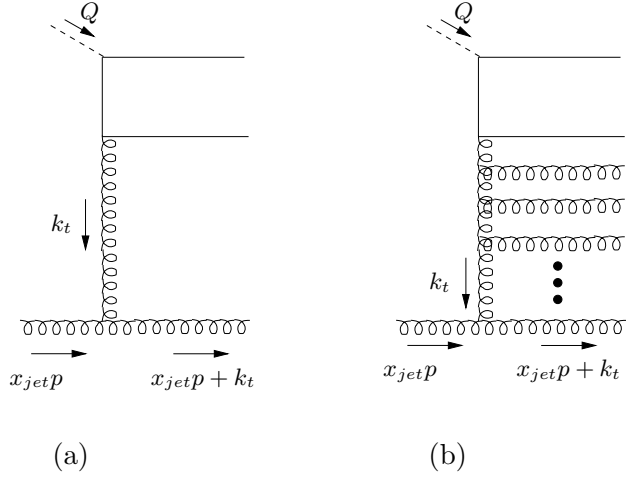}
}
\end{picture}
\end{center}
\caption[]{\label{fig1}                                                        
\sl Generic diagrams for the process (a) $g \to g  +                           
q\bar{q}$ and (b) $g \to g + \,(n\, g)\, +                                     
q\bar{q}$. Diagrams with an incident quark (not shown) have been               
considered as well.}                                                           
\end{figure}  
                                                                              
%
Using the gauge condition $(q+xp) \cdot A =0$, where $q$ and $p$ are the       
four-momenta of the virtual photon and proton, respectively, the                
 diagrams                                                                      
corresponding to the two processes                                             
$g \rightarrow gq\bar{q}$ and $q \rightarrow qq\bar{q}$                    
have to be calculated. The incoming parton is always scattered into the        
forward region, well separated in rapidity from the photon region.             
Moreover, the two diagrams differ only in their colour content, since          
the coupling of the $t$-channel gluon to the incoming parton (gluon or
quark) is of eikonal type. Due to the large rapidity gap                       
given by $\ln(x_{jet}/x_B)$ 
 the probability to emit more                                                  
gluons is increased, for the smallness of the strong coupling constant         
$\alpha_s$ is compensated by the large logarithm $\ln(x_{jet}/x_B)$.
This means                                                                     
that higher-order corrections become relevant, and in the end                  
all diagrams with an arbitrary number of $s$-channel gluons have to be 
considered. A generic diagram of this process is shown in Fig.~1b.             
Their resummation is performed in terms of the BFKL equation                   
\cite{BFKL} and leads to the following result for the one-parton inclusive     
cross section \cite{Hot}:                                                     
\beqn\label{e1}                                                                
\frac{d \sigma}{dx_{jet}                                                       
 dk^2_t dy dQ^2}&=&                                                            
\frac{\alpha^2_{em}}{2yQ^4}\;\left\{[1+(1-y)^2]\;x_BW_t^{(0)}       
\;+\;4(1-y)\;x_BW_l^{(0)}\right\} \nonumber \\                                
&&\cdot\; \frac{3\alpha_s}{\pi k_t^4}\;\left\{g(x_{jet},                       
k_t^2)\;+\;                                                                    
\frac{4}{9}\;\sum_f q_f(x_{jet},                                               
k_t^2)\right\}
\eeqn                                                                          
with                                                                           
\beqn\label{e2}                                                                
x_BW_t^{(n)}&=&\sum_f Q^2_f\;\alpha_s \;\sqrt{k_t^2Q^2}\;\frac{\pi}{4}\;
\int_0^\infty d\mu \;\frac{\frac{9}{4}+\mu^2}{1+\mu^2}\;\frac{\sinh(\mu\pi)}   
{\mu\cosh^2(\mu\pi)} \;\;\nonumber\\                                      
&&\spc{3cm}\cdot\;\;\left(\frac{x_{jet}}                                       
{x_B}\right)^{\omega (n,\mu)}\;                                         
\cos\left(\mu\ln(\frac{k_t^2}{Q^2})\right) \;\; ,                              
\eeqn                                                                          
\beqn\label{e3}                                                                
x_BW_l^{(n)}&=&\sum_f Q^2_f\;\alpha_s \;\sqrt{k_t^2Q^2}\;\frac{\pi}{4}\;
\int_0^\infty d\mu \;\frac{\frac{1}{4}+\mu^2}{1+\mu^2}\;\frac{\sinh(\mu\pi)}   
{\mu\cosh^2(\mu\pi)} \;\;\nonumber\\                                      
&&\spc{3cm}\cdot\;\;\left(\frac{x_{jet}}                                       
{x_B}\right)^{\omega (n,\mu)}\;                                         
\cos\left(\mu\ln(\frac{k_t^2}{Q^2})\right)                                     
\eeqn                                                                          
where                                                                         
\be \label{e4}                                                                 
\omega (n,\mu)\;=\;\frac{3\alpha_s}{\pi}\left\{2\psi(1)\;-\;                   
\psi\left(\frac{n+1}{2}+i\mu\right)\;-\;\psi\left(\frac{n+1}{2}-i\mu\right)
\right\}\;\;.             
\ee                                                                            
Here $n\ge 0$ is an integer, and                           
$W_t$ and $W_l$ refer to the transverse and the                                
longitudinal parts of the hadronic                                             
tensor $W^{\rho\sigma}$. The transverse momentum of the jet                    
and its longitudinal                                                           
energy fraction are given by $k_t$ and $x_{jet}$.                              
The first-order running coupling constant was used with $k_t$ as scale.    
The main features of the BFKL prediction, in particular the power-like         
increase at small $x_B$, can immediately be read off
from Eqs.~(\ref{e2}) and (\ref{e3}), making use                                
of the fact that the integrand dominates at $\mu=0$. For $\mu=0$ and $n=0$     
one finds 9/2 for the ratio of the transverse and the longitudinal part of     
the cross section. For large $\ln(x_{jet}/x_B)$ 
the BFKL power is obtained from                
$\omega (0,0)$. Before we discuss our numerical results based upon             
this BFKL formula, we briefly turn to the fixed-order matrix elements.\\ \\    
{\bf 4.~Fixed-Order Matrix Elements Calculations:}                             
In order to compare the BFKL results with a fixed-order 
calculation based on exact matrix elements, we study the contributions   
for the production of two and three partons on the Born level,                 
giving rise to (2+1)- and (3+1)-jet events.                                     
We use the PROJET Monte Carlo program \cite{Gra94}, where the cross         
sections from Ref.~\cite{GM91} are implemented
in the modified JADE jet-definition scheme\footnote{Another calculation of
(3+1)-jet cross sections can be found in Ref.~\cite{BKMS89}.
It should be noted that the JADE jet-definition scheme is different
from the cone scheme employed in the experimental analysis \cite{H1}}. 
Two partons can be produced in the parton-level processes                      
$q\rightarrow qg$ and $g\rightarrow q\overline{q}$;                            
for three partons                                                              
there are the processes                                                        
$q\rightarrow qgg$,                                                            
$g\rightarrow q\overline{q}g$                                                  
and                                                                            
$q\rightarrow qq\overline{q}$.                                                 
These matrix elements suffer from infrared and collinear singularities,        
which are excluded by means of suitable cuts on the parton momenta.            
This is done by requiring that the invariant mass squared                      
$s_{ab}=2p_a p_b$ be larger than some mass scale $M_{cut}^2 = c W^2$, $c$
being the jet cut parameter and $W^2$ the 
squared invariant mass of the hadronic final state;                  
$a$ and $b$ are set to $q$, $\overline{q}$, $g$ and $r$, where $p_r$ is        
the momentum of the proton remnant jet.                                        
All matrix elements except those from the class $g\rightarrow q\overline{q}g$  
are singular                                                                   
if any of the invariants goes to zero. The matrix element                      
$g\rightarrow q\overline{q}g$                                                  
is not singular for $s_{q\overline{q}}=0$, because of the additional           
propagator of the quark line.                                                  
In the following numerical study, we have used the same cuts for the           
forward jet as those that are applied                                          
by the H1 collaboration in their BFKL study. The parton density used is the    
CTEQ3 leading-order parametrization.\\ \\                                      
{\bf 5.~Numerical Results:}                                                    
Numerical results for both approaches are shown in Fig.~2.                     
Let us begin with the BFKL approach. The plots depicted in Fig.~2a were        
evaluated                                                                      
by integrating the differential cross section within the given cuts            
(see above). The forward jet selection A is used.                              
Two cases are considered:\\
\begin{itemize}
\item[(i)] The fixed-order but high energy                   
asymptotic expression for the two processes $q \to q +q\bar{q}$ and            
$g \to g + q\bar{q}$ (lower two curves in Fig.~2a). These                      
cross sections are evaluated in the BFKL-type                                  
high energy approximation, and from Eqs.~(\ref{e2}) and (\ref{e3}) they are
obtained by simply setting $\omega (n,\mu) = 0$.\\
\item[(ii)] The full BFKL result 
according to Eqs.~(\ref{e2}) and (\ref{e3}) (the upper 
two curves in Fig.~2a). Only light quarks $(u, d, $ and $s$) are 
included.
The curves were scaled by 20\% to approximately include the effect of 
charm production, in accord with recent data \cite{deroeck}.
\end{itemize}
Case (i) shows                                          
good agreement with the results of the second approach, the                    
full matrix element calculations (Fig.~2b, see below).                         
The full BFKL approach, however,                                               
scales up the normalization roughly by a factor                                
$(x_{jet}/x_B)^{0.5}$ 
and reveals a much steeper shape in the $x_B$-distribution.
In comparison with data,                    
it is in particular the shape of the cross section which is an important
test for the presence of BFKL. The absolute normalization suffers from 
some uncertainty as discussed below.
Figure~2c 
shows the same calculation for the second set of kinematical cuts.

\begin{figure}[thbp] \unitlength 1mm
\begin{center}
\dgpicture{159}{165}
 
 \put( 10, 90){\epsfigdg{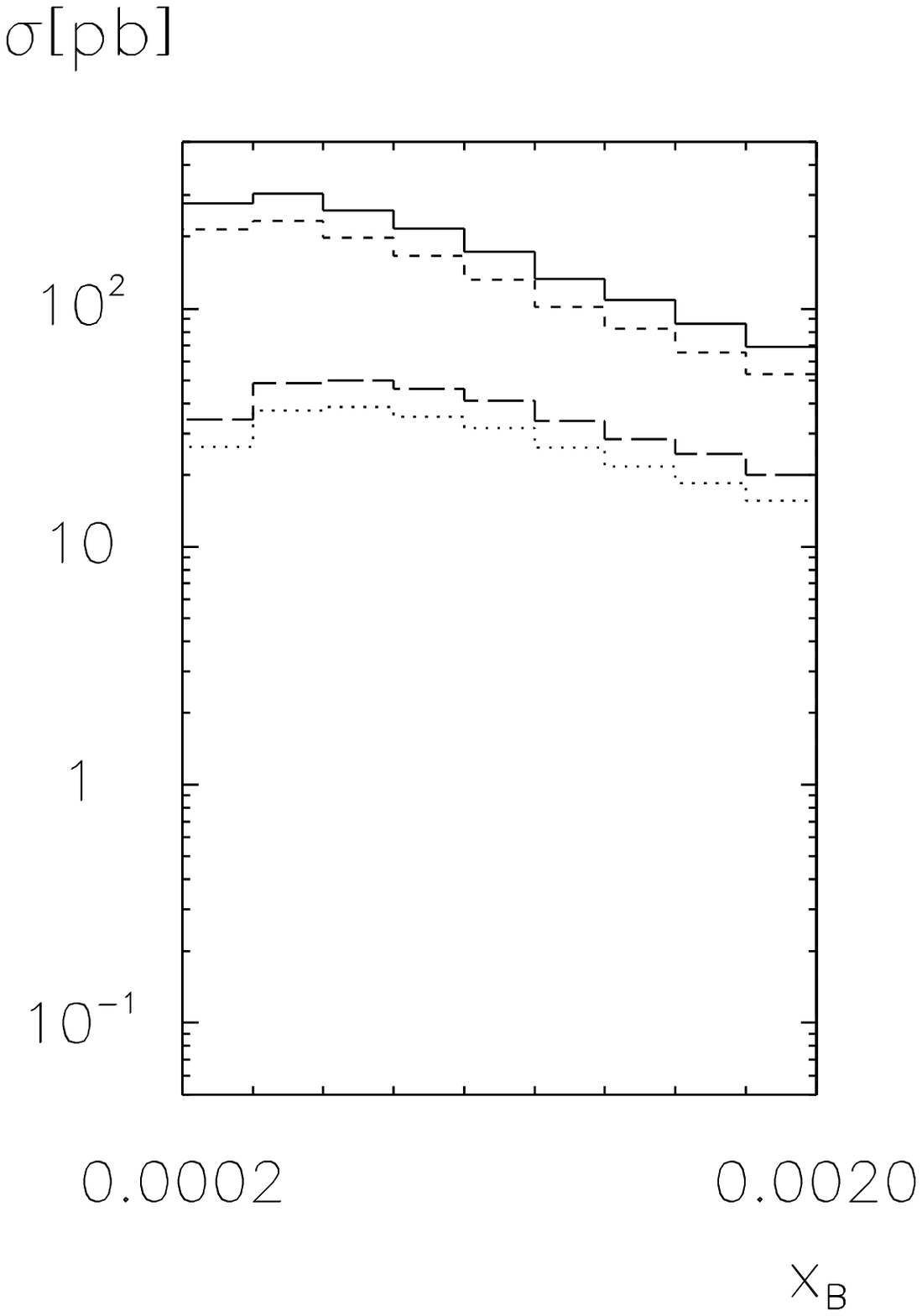}{width=60mm}}
 \put( 56,160){\rm (a)}   
  
 \put( 90, 90){\epsfigdg{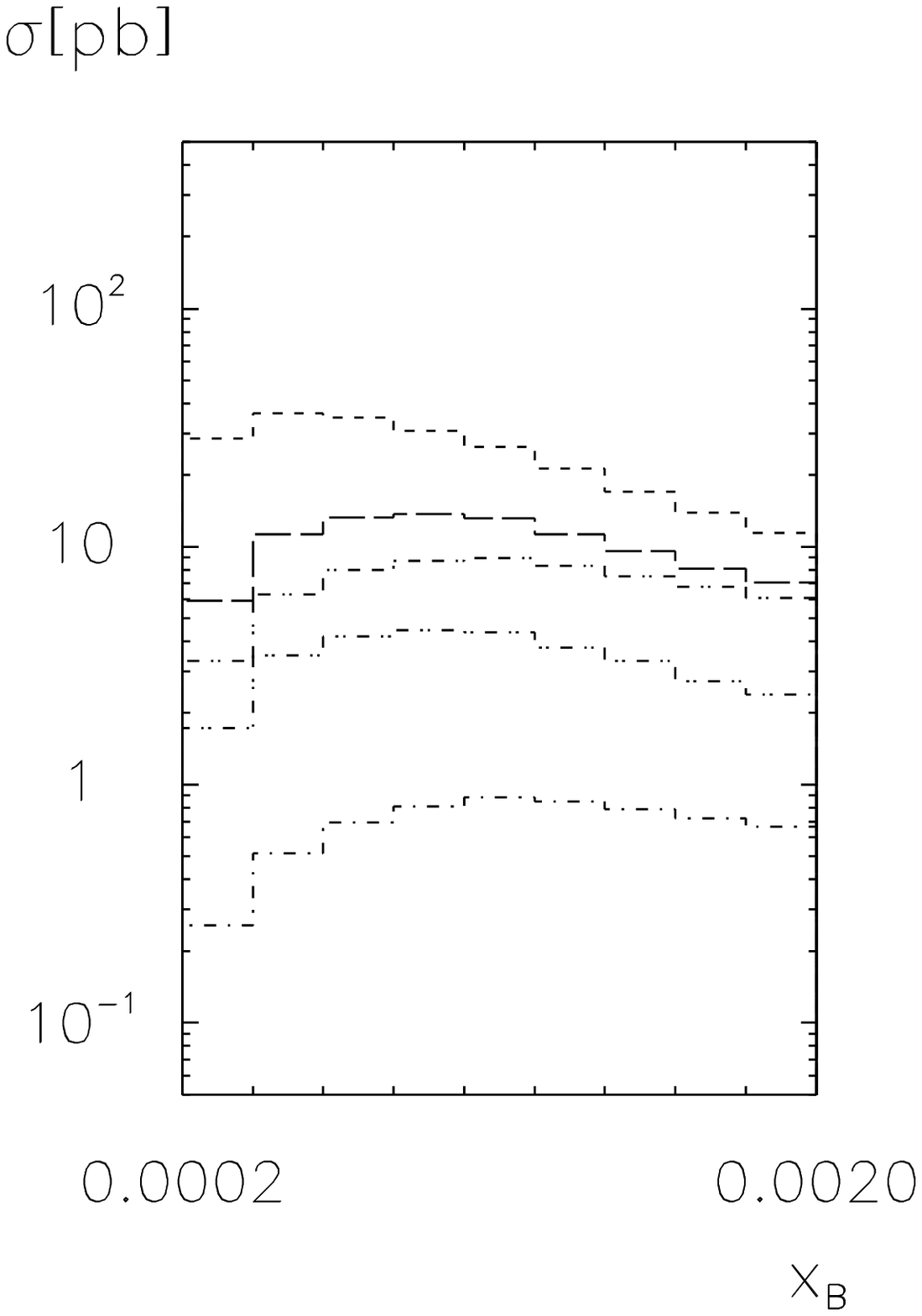}{width=60mm}}
 \put(136,160){\rm (b)}   
   
 \put( 10, 0){\epsfigdg{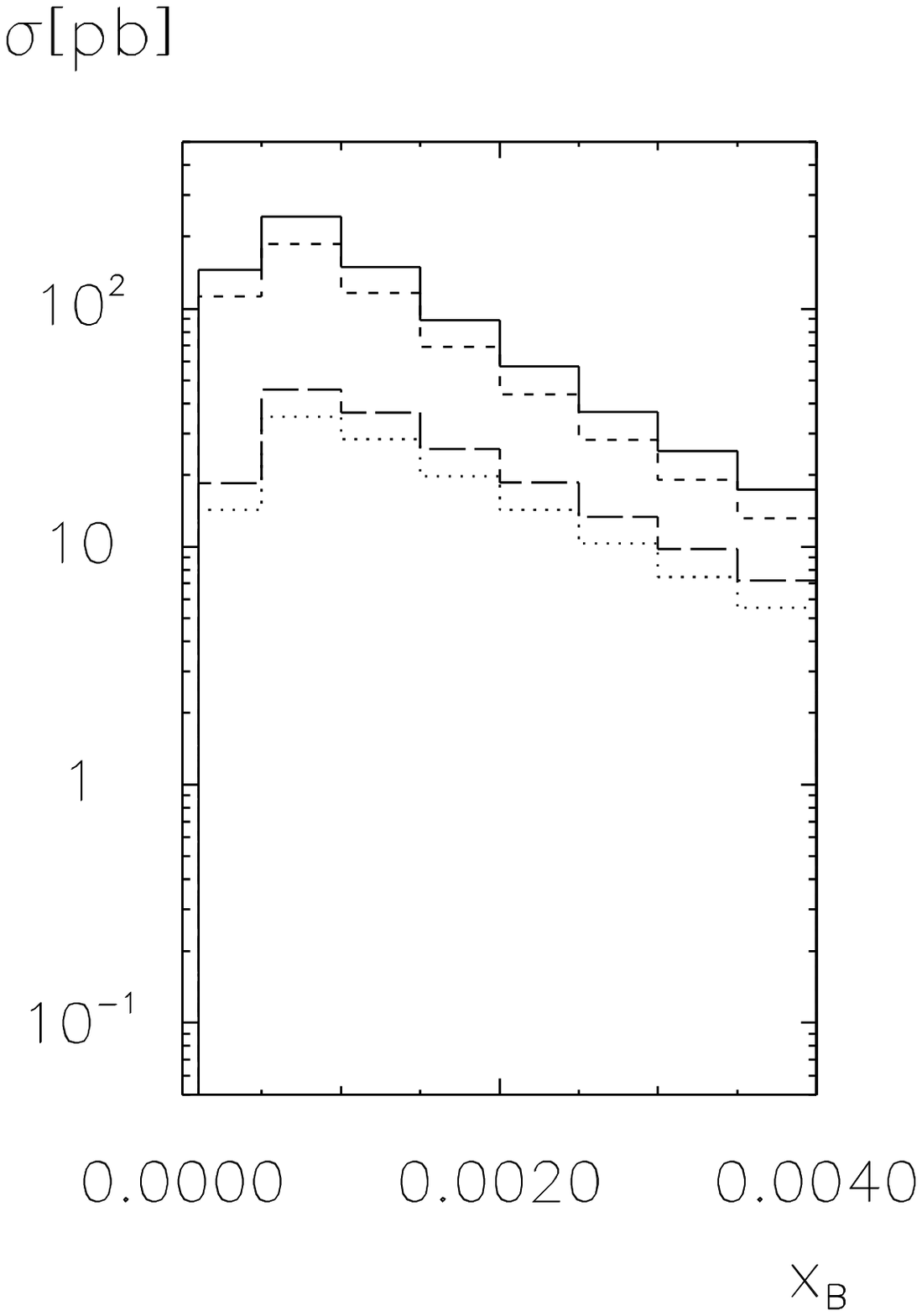}{width=60mm}}
 \put( 56,70){\rm (c)}
 
 \put( 90, 0){\epsfigdg{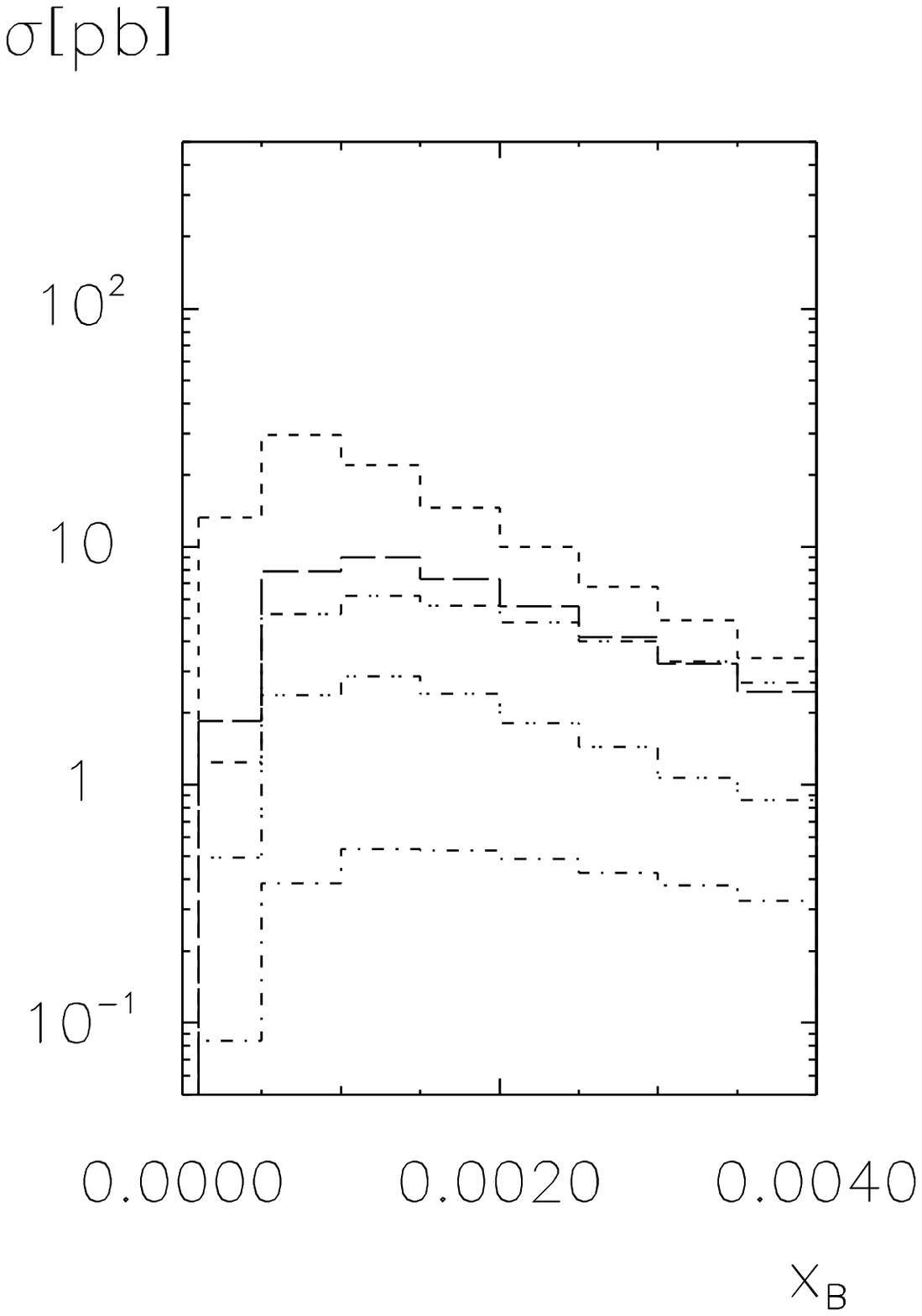}{width=60mm}}
 \put(136,70){\rm (d)}
  
\end{picture}
\end{center}
\caption[]{\label{fig2}                                                        
\sl (a) $x_B$ dependence of the cross sections                                 
from the BFKL calculation                                                      
(incoming quark and gluon \mbox{(\fullline)}                                   
and incoming gluon only \mbox{(\dashline)})                                    
and from an approximate analytical calculation                                 
of the three-parton matrix elements                                            
(incoming quark and gluon \mbox{(\longdashline)}                               
and incoming gluon only \mbox{(\dotline)}).                                    
(b) $x_B$ dependence of the cross sections from the exact fixed-order          
matrix elements, with any of the partons as the forward jet.                   
$g\rightarrow q\overline{q}g$                                                  
integrated over the full range of the invariant $s_{q\overline{q}}$            
\mbox{(\dashline)};                                                            
the same, but with the additional invariant mass cut                           
$s_{q\overline{q}}>M_{cut}^2$                                                  
\mbox{(\longdashline)};                                                        
$q\rightarrow qgg$ \mbox{(\dashdotline)} and $q\rightarrow qq\overline{q}$     
\mbox{(\dashdotdotline)},                                                      
with                                                                           
the cut $s_{AB}>M_{cut}^2$ on all                                              
invariants;                                                                    
sum of $q\rightarrow qg$ and $g\rightarrow q\overline{q}$                      
\mbox{(\dashdashdotdotline)},                                                  
with the cut $s_{AB}>M_{cut}^2$ on all invariants.                             
The jet cut parameter is  $c=0.005$.  The distributions              
(c) and (d) are the same as (a) and (b), but for the jet cuts set B
and calculated for $10^{-4}<x<4\times 10^{-3}$.}            
\end{figure}                                                                  

\begin{figure}[thbb] \unitlength 1mm
\begin{center}
\dgpicture{159}{82}
 
\put( 30,0){\epsfigdg{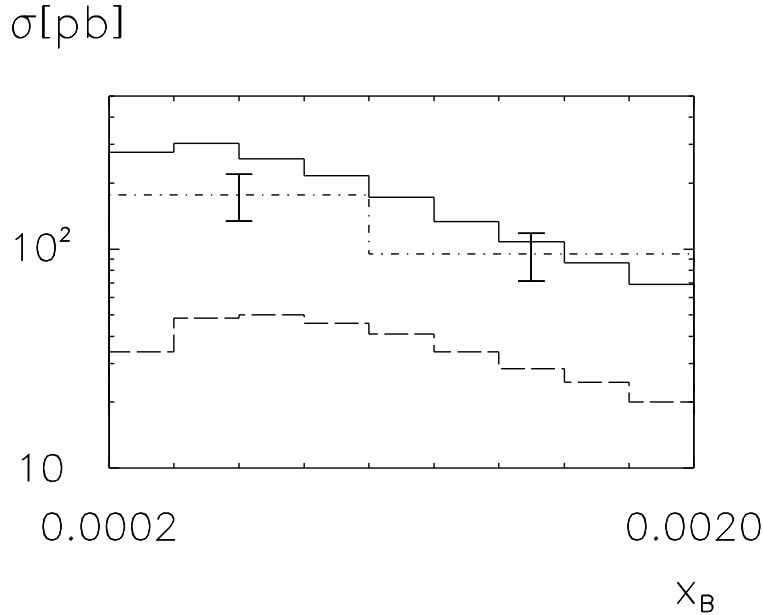}{width=100mm}}
  
\end{picture}
\end{center}
\caption[]{\sl
Comparison of the BFKL calculation and the approximate analytical 
calculation of the three-parton matrix elements, shown in Fig.~2a, 
compared with the data of the H1 experiment
\protect \cite{H1}.\label{fig3}}
\end{figure}

Next we turn to the                                                            
second approach, the fixed-order matrix element calculation.               
The results for the $x_B$-dependence are shown in Figs.~2b
and 2d for the jet selections A and B, respectively. As a result of the    
applied cut, the contribution from the (2+1)-jet matrix element is small.      
The (3+1)-jet contribution $q \to qgg$ is small as well; the quark densities   
are small compared with the gluon densities, 
and the process is not a $t$-channel
process. The $q \to qq\bar{q}$ 
contribution is larger, because in this class of diagrams    
there is a $t$-channel 
gluon which enhances the cross section for forward-jet
production.
The (3+1)-jet cross section is dominated by the process $g \to gq\bar{q}$.     
The gluon density is large, and there are large contributions from the         
diagrams with a $t$-channel gluon. In order to compare the full fixed-order    
matrix element to the BFKL prediction, it is, in principle, necessary to       
calculate a cross section that is inclusive in all particle momenta            
except for the momentum of the forward jet. In the present version of PROJET,  
such a study is not feasible. However, it is possible to remove the cut        
$s_{q\bar{q}}>M_{cut}^2$ for the process $g \to g q\bar{q}$ (see the 
above discussion). 
Relaxing this cut increases the cross section considerably. The shape  
of the $x_B$ distribution compares well with that of the approximate analytic
calculation for the process $g \to gq\bar{q}$ (see above), whereas for the     
full BFKL result (Fig.~2a) the slope in $x_B$ is somewhat steeper.           
It should be noted that                                                        
the cross section from the exact matrix elements are calculated from a         
modified JADE scheme and are less ``inclusive'' than those from the            
approximate analytic calculation; a direct comparison of the magnitude is      
therefore not possible.                                                        
\\ \\                                                                          
Finally, we compare the analytic predictions with the H1  data. 
Because of the JADE jet definition scheme, the fixed-order matrix elements 
are not quantitatively comparable with the data based on the cone scheme. 
Moreover PROJET treats the production of charm quarks in the massless
approximation.
The analysis 
presented in Ref.~\cite{H1}
 has used only two $x_B$-bins, and the two values are     
shown in Fig.~3. 
Obviously, the data are higher than the analytic fixed-order  
results, indicating that gluon production, which is responsible for            
the rise of the BFKL Pomeron, is present. On the                               
other hand, the data points lie very close to  the BFKL prediction.            
This looks very encouraging and could be interpreted as the first sign         
of the presence of the BFKL Pomeron at HERA.                                   
However, we feel that several remarks are in order. 
First, for the analogous process at hadron colliders (Mueller-Navelet
jets) a comparison of a next-to-leading order QCD matrixelement
calculation with the BFKL approximation has shown \cite{D2}
that the latter one
overestimates the available phase space. A more accurate treatment of
the forward jet production, therefore, would most likely tend to
lower the BFKL prediction.
A similar effect has to be expected also in      
the forward-jet production at HERA. 
Secondly, our treatment of the charm       
contribution is clearly unsatisfactory, and 
a more accurate treatment of the charm mass has to be included 
in our calculation.
Finally, as we have mentioned in the beginning,           
also hadronization which is not included in any of our calculations may        
have some effect. We therefore feel that we have to be cautious in             
drawing too strong conclusions. It seems safe to say that - within the         
accuracy to which the data can be compared with our calculations -             
the measurement of forward jets does not agree with the 
approximate fixed-order matrix     
element calculation, and                                                       
the comparison with the BFKL prediction is very encouraging.                   
However, there may be some effects, which can distort this good agreement.      
\\ \\                                                                          
\begin{figure}[thbp] \unitlength 1mm
\begin{center}
\dgpicture{159}{165}
 
 \put( 10, 90){\epsfigdg{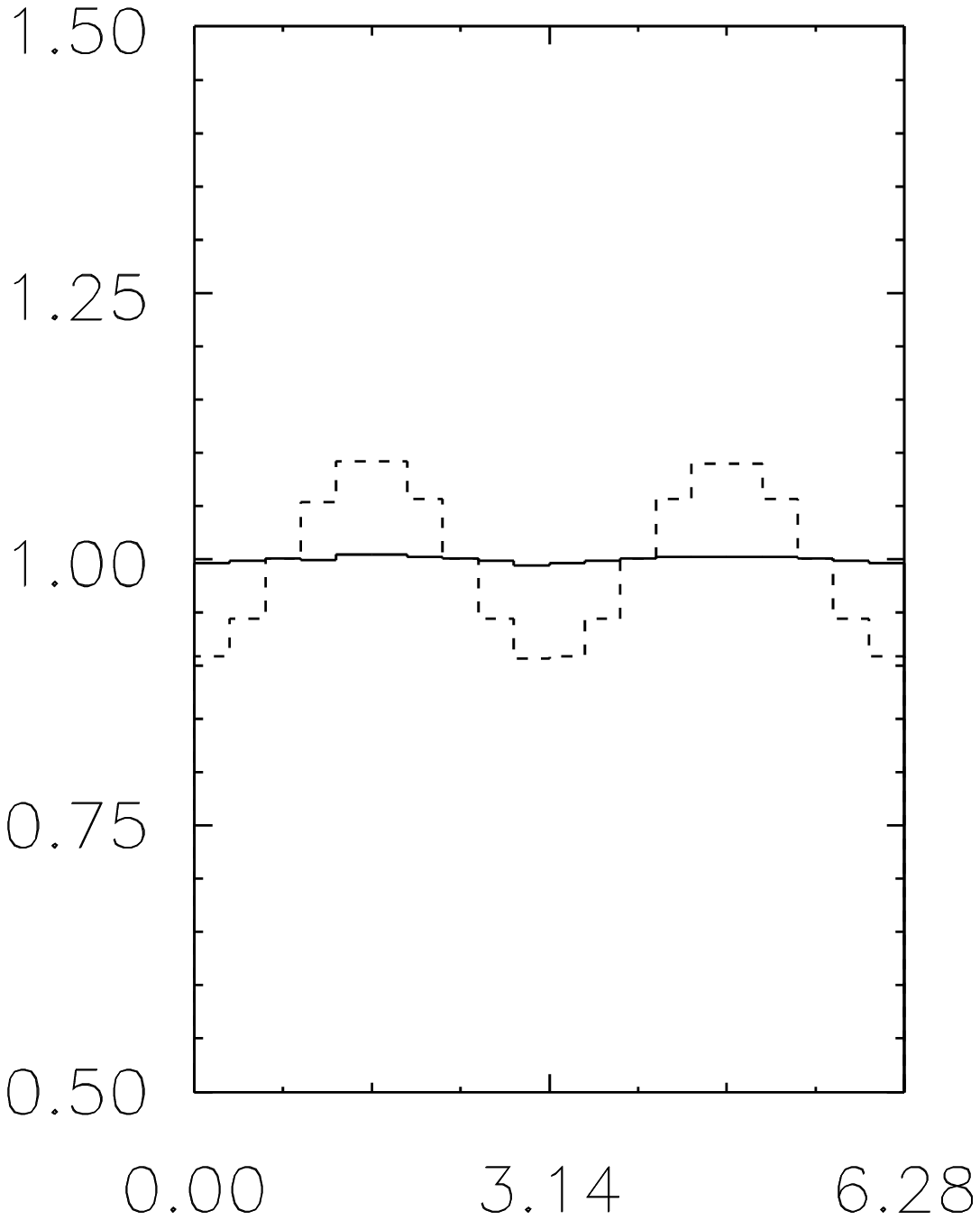}{width=60mm}}
 \put( 58,158){\rm (a)}   
  
 \put( 90, 90){\epsfigdg{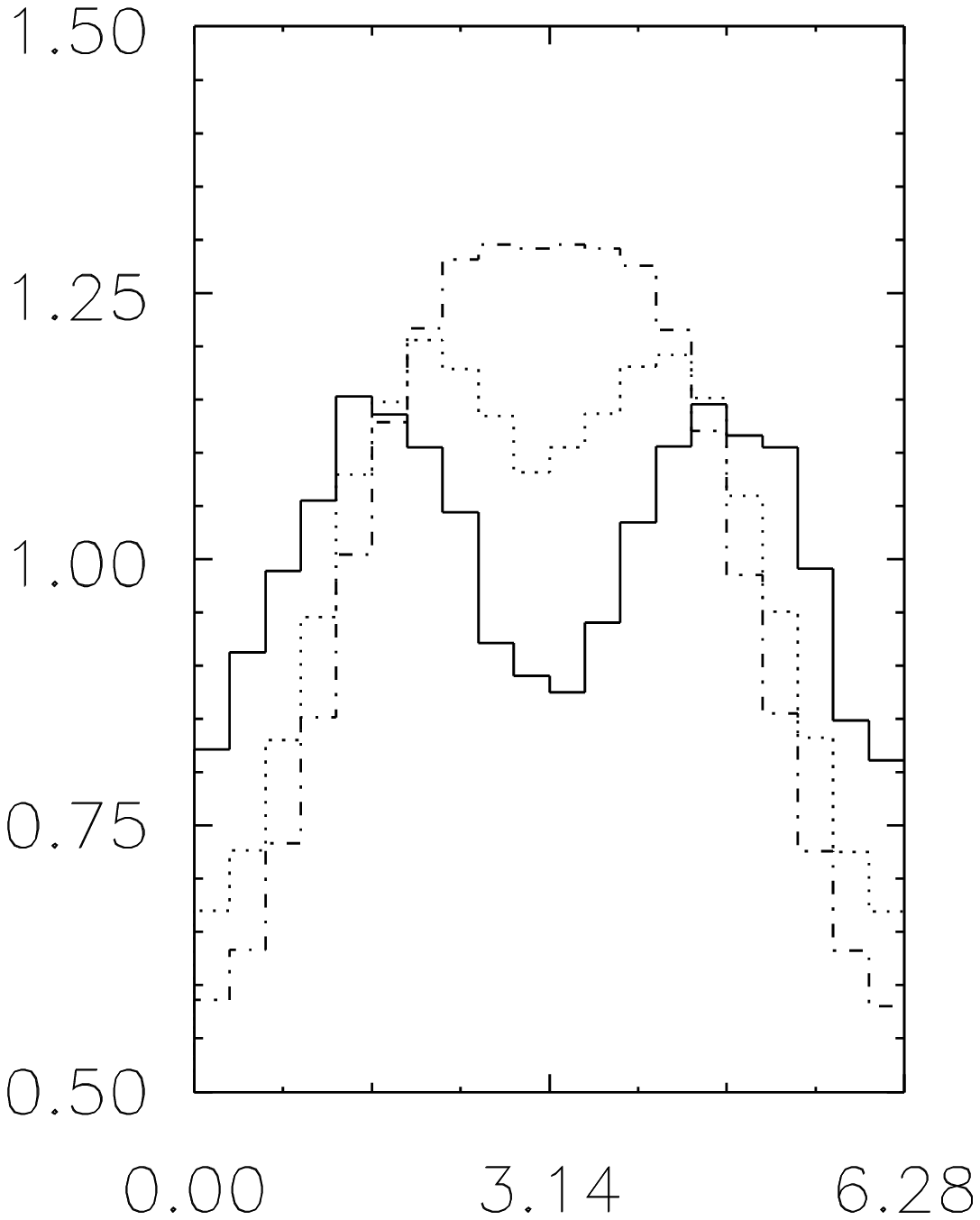}{width=60mm}}
 \put(138,158){\rm (b)}   
   
 \put( 10, 0){\epsfigdg{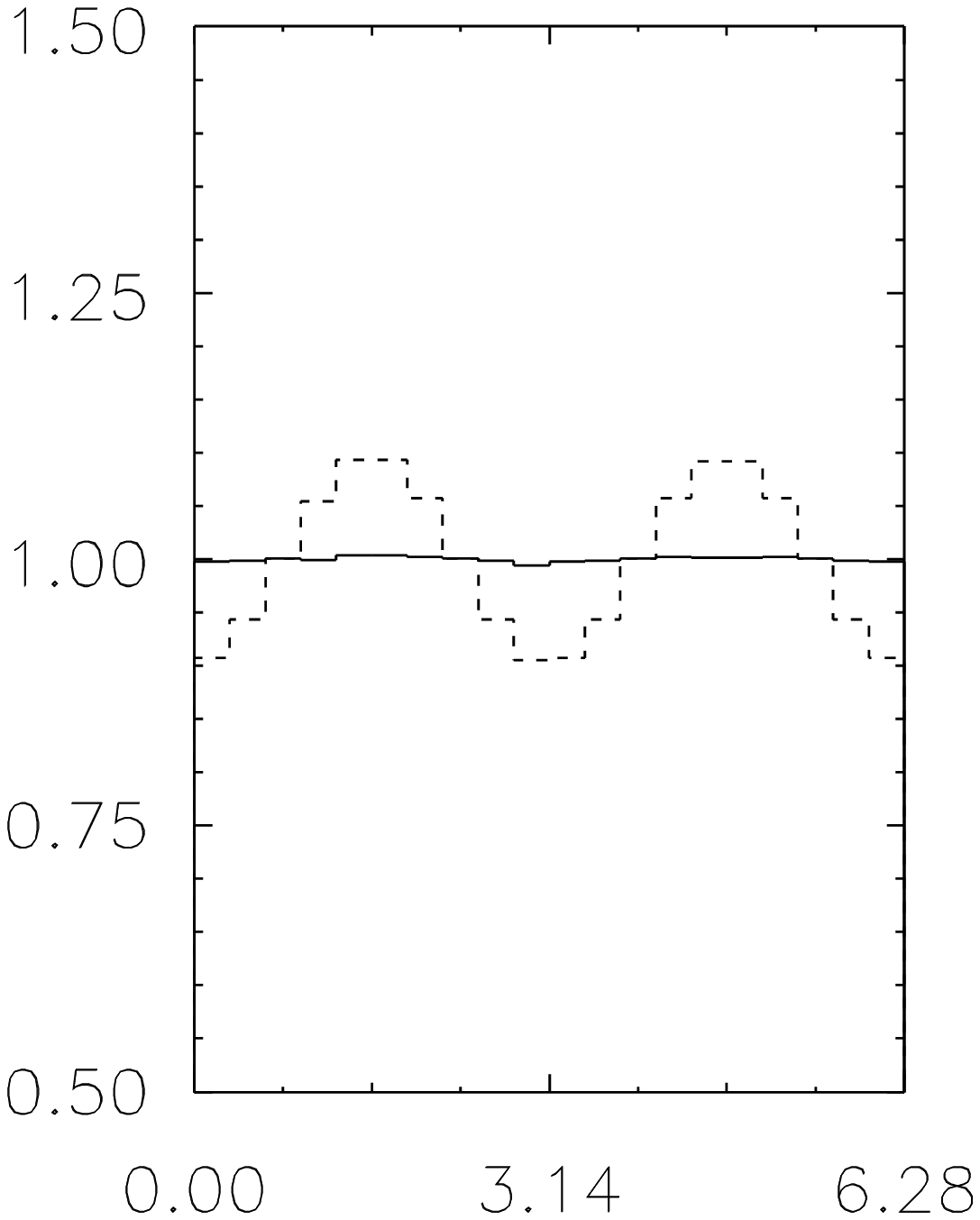}{width=60mm}}
 \put( 58,68){\rm (c)}
 
 \put( 90, 0){\epsfigdg{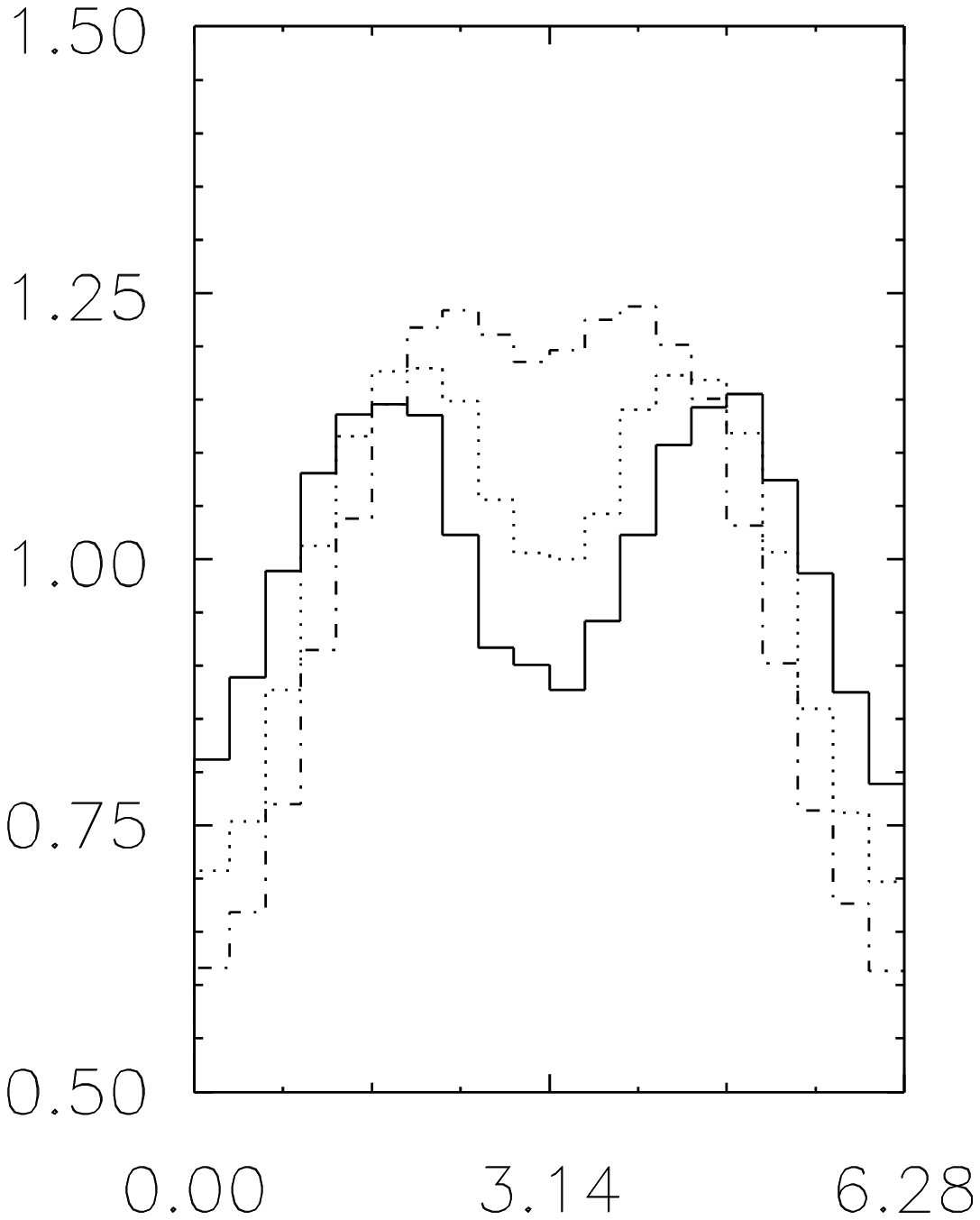}{width=60mm}}
 \put(138,68){\rm (d)}
  
\end{picture}
\end{center}
\caption[]{\label{fig4}                                                        
\sl
Dependence
on the difference $\Phi$ of the azimuthal angles of the                        
outgoing lepton and the forward jet,                                           
normalized to a common average.                                                
(a)                                                                         
BFKL result, the bin in $x_B$ is                                               
$[0.2\times10^{-3},0.4\times10^{-3}]$ \mbox{(\fullline)}.
Approximative analytical Born result, the bin in $x_B$ is                      
$[0.2\times10^{-3},0.4\times10^{-3}]$ \mbox{(\dashline)}.
(b) 
Fixed-order result, 
the bins in $x_B$ are given by                                                 
$[0.2\times10^{-3},0.4\times10^{-3}]$ \mbox{(\fullline)},
$[1.8\times10^{-3},2.0\times10^{-3}]$ \mbox{(\dotline)} and         
$[3.4\times10^{-3},3.6\times10^{-3}]$ \mbox{(\dashdotline)}.        
The jet cut parameter  $c=0.005$.                
The matrix element is                                                          
$g\rightarrow q\overline{q}g$,                                                 
integrated over the full range of the invariant $s_{q\overline{q}}$,           
where any of the partons may be the forward jet.  
The distributions                             
(c) and (d) are the same as (a) and (b), but for the jet cuts set B.}              
\end{figure}                                                                  

\noindent                                                                      
{\bf 6.~Angular decorrelation:}                                                
Recently \cite{BDW} the azimuthal dependence of the jet production            
cross section has been calculated\footnote{A somewhat different type of       
azimuthal dependence has been studied in Ref.~\cite{AKMS}.}. The angle 
$\Phi$ is defined in the HERA                                                  
frame as the angle between the transverse momenta of the outgoing lepton       
and the forward jet, and the formula for the cross section reads:              
\beqn\label{e5}                                                                
\frac{d \sigma}{dx_{jet}                                                      
 dk^2_t d\Phi dy dQ^2}&=&                                                      
\frac{\alpha^2_{em}}{2yQ^4}\;\left\{[1+(1-y)^2]\;x_BW_t^{(0)}
\;+\;4(1-y)\;x_BW_l^{(0)}\right. \nonumber\\                                   
&&\left.\spc{2cm}-\; 2(1-y)\;\cos(2\Phi)\;x_BW_l^{(2)}\right\}\\  
&&\cdot\; \frac{3\alpha_s}{2\pi^2 k_t^4}\;\left\{g(x_{jet},                    
k_t^2)\;+\;                                                                    
\frac{4}{9}\;\sum_f q_f(x_{jet},                                               
k_t^2)\right\}\nonumber\;\;.                                                   
\eeqn                                                                          
$W_l^{(2)}$ is given by Eq.~(\ref{e3}) with $n=2$.                              
For small $1/x_B$, Eq.~(\ref{e5}) predicts a maximum of the cross section
at $\Phi=\pi/2$. When $1/x_B$ gets large, we expect                      
in the BFKL prediction the $\Phi$-dependence to disappear. In Eq.~(\ref{e5})    
this effect follows from the difference of the powers                          
$\omega (0,0)\simeq 0.50$ and $\omega(2,0) \simeq -0.22$.                      
With increasing                                                                
$x_{jet}/x_B$, the angular dependent part  
is suppressed whereas all other parts increase.\\ \\                     
Numerical results are shown in Fig.~4.                                          
In the BFKL case (Fig.~4a) the                                                 
$\Phi$-dependence is completely washed out when $x_B$ becomes small 
(horizontal  
line in Fig.~4a). This is                                                      
in complete agreement with the theoretical expectation. For the 
approximate fixed-order process        
$g \to g q\bar{q}$, on the other hand,    
the cross section is not flat. It has a clear maximum at $\Phi=\pi/2$
and is found to be rather insensitive to the choice of the $x_B$ bin
(not shown).   
Figure~4c 
shows the same calculations as Fig.~4a, for the jet selection set B.     
\\ \\                                                                          
For comparison, we                                                             
also show the results of the fixed-order matrix element calculation            
(Fig.~4b).                                                                      
We have used the jet cut $M_{cut}^2=c W^2$ with $c=0.005$.  The curve      
shown in Fig.~4b 
belongs to the process $g \to g q\bar{q}$, integrated over the 
full range of the invariant $s_{q\bar{q}}$. The forward jet may belong to any  
of the partons. At smaller $x_B$, there is a tendency to enhance the maximum
of the distribution at $\Phi=\pi/2$ and $\Phi=3\pi/2$, similar to the          
BFKL Born calculation in Fig.~4a. There is a clear                             
difference with the full BFKL prediction in Fig.~4a.
At larger $x_B$,
on the other hand, the full matrix element calculation does not agree very    
well with the BFKL Born calculation.                                          
Figures 4c and 4d 
show the results for selection B of the jet cuts.\\ \\                
The present detectors at HERA allow for jet angular measurements with a        
resolution of about $5^\circ$--$10^\circ$ 
for jets with an $E_t$ larger than 5 GeV.                                      
This opens a window for measuring angular correlations between             
struck quark and forward jet, or between scattered electron and        
forward jets. The major background for this measurement is made up of 
radiative events; due to the changed kinematics at                       
the hadron vertex,
these can eject a jet in the forward direction.                    
This                                                                           
background is however expected to be controlled experimentally to a sufficient 
level to allow for a measurement of better than
a  10\% significance level in large       
data samples. We therefore expect that a measurement of the angular            
decorrelation and the comparison with theoretical expectations                 
provides a further test of BFKL dynamics.                                      
\\ \\                                                                          
\noindent                                                                      
{\bf 7.~Conclusions:}                                                          
In this paper we have made a comparison for the production cross section of   
associated jets in the forward direction of the BFKL prediction and of        
the fixed-order matrix element calculation. Both curves show a clear           
difference at small $x_B$,
thus confirming the expectation that this process       
provides a clean signal of the BFKL Pomeron. Confrontation with the data       
is encouraging: the data clearly lie above the fixed-order calculation         
and prove that the radiation of extra gluons is an important effect.           
The BFKL calculation is very close to the data, which could be the first sign  
of the presence of the BFKL Pomeron in the                                     
HERA data.                                                                     
However, there are several aspects that need further study, in particular     
the contribution of charm and hadronization effects in the final state.        
{}From 
the theoretical side, there is also the question of 
the kinematic    
region in which
the BFKL approximation is applicable. We expect that the strongest        
correction will come from unitarization effects, and we know that first        
unitarity corrections will lower the BFKL prediction.                          
Hence it may be premature to interpret the data as a proof of          
the presence of the BFKL Pomeron.\\ \\                                         
For future measurements we have suggested to look also into the azimuthal      
dependence of the jet cross section. Comparison with fixed-order calculations  
show that this provides a new signal for BFKL dynamics, which may help to       
find further footprints of the BFKL Pomeron at HERA.                           


\begin{thebibliography}{xx}                                                    
\bibitem{BFKL} E.A.~Kuraev, L.N.~Lipatov and V.S.~Fadin,              
{\em Sov.\ Phys.\ JETP} {\bf 45} (1977) 199; \\
Ya.Ya.~Balitskij and L.N.~Lipatov, {\em Sov.\ J.~Nucl.\ Phys.}              
{\bf 28} (1978) 822.                                                           
\bibitem{Mue} A.H.~Mueller, {\em Nucl.\ Phys.~B} (Proc.\ Suppl.) {\bf 18C}
(1991) 125.                                                                    
\bibitem{Hot} J.~Bartels, A.~De Roeck and M.~Loewe, 
{\em Z.~Phys.} {\bf C54}         
(1992) 635;\\                                                                  
J.~Kwiecinski, A.D.~Martin and P.J.~Sutton, {\em Phys.\ Lett.} 
{\bf B287} (1992) 254; 
{\em Phys.\ Rev.} {\bf D46} (1992) 921; \\                               
W.-K.~Tang, {\em Phys.\ Lett.} {\bf B278} (1991) 363;\\                   
J.~Bartels, M.~Besan\c{c}on, A.~De Roeck and J.~Kurzhoefer, in               
{\em Proceedings of the HERA Workshop 1992} (eds.\ W.~Buchm\"uller and    
G.~Ingelman), p.~203.                                                     
\bibitem{H1} H1 Collab., DESY-95-108 and {\em Phys.\ 
Lett.} {\bf B356} (1995) 118.  
\bibitem{BDW}J.~Bartels, V.~Del Duca and M.~Wuesthoff, 
DESY preprint in preparation.
\bibitem{MN} A.H.~Mueller and H.~Navelet, 
{\em Nucl.\ Phys.} {\bf B282} (1987) 727.  
\bibitem{D} V.~Del Duca and C.R.~Schmidt, {\em Phys.\ Rev.} 
{\bf D49} (1994) 4510;\\
W.J.~Stirling, {\em Nucl.\ Phys.} {\bf B423} (1994) 56.
\bibitem{GM91} D.~Graudenz and N.~Magnussen, in 
{\em Proceedings of the HERA        
Workshop 1991}, DESY (eds.\ W.~Buchm\"uller and G.~Ingelman).
\bibitem{Gra94} D.~Graudenz, PROJET 4.13 manual, 
{\em Comput.\ Phys.\ Commun.} {\bf 92} (1995) 65.
\bibitem{BKMS89} T.~Brodkorb, J.G.~K\"orner, E.~Mirkes and G.A.~Schuler,
{\em Z.~Phys.} {\bf C44} (1989) 415.                                         
\bibitem{deroeck} A.~De Roeck, to appear in {\em Proceedings of the 2nd
Epiphany
Workshop}, Krakow, 1996. 
\bibitem{D2} V.~Del Duca and C.R.~Schmidt, {\em Phys.\ Rev.} 
{\bf D51} (1995) 2150.  
\bibitem{AKMS}A.~Askew, D.~Graudenz, J.~Kwiecinski and A.D.~Martin,  
{\em Phys. Lett.} {\bf B338} (1994) 92.
\end{thebibliography}
\end{document}